\documentclass[12pt]{article}
\usepackage{graphics}
\begin{document}
\begin{center}
{The Non-perturbative term for the Axial-vector Form Factor of Pion Decay}
\end{center}
\begin{center}
{Susumu Kinpara}
\end{center}
\begin{center}
{\it Institute for Quantum Medical Science (QST)\\ Chiba 263-8555, Japan}
\end{center}
\begin{abstract} 
The axial-vector form factor for the decay of pion $\pi^{+} \rightarrow \gamma + e^{+} + \nu_e$ is calculated 
by using the pseudovector coupling pion-nucleon interaction with the non-perturbative term.
The self-energy is approximated by the lowest-order constant.
The parameter is significant to improve the value of the form factor as well as the vector form factor in our previous study.
\end{abstract}
\section*{\normalsize{1 \quad Introduction}}
\hspace*{4.mm}
The study of the pion-nucleon interacting system has been continuing.
The results of the calculation for the process of the scattering
have revealed that the pion plays the central role to describe the various phenomena.
It is different from the nuclear many-body system where the heavier mesons such as the sigma and the omega are important to construct the structure of nuclei
since the Hartree approximation of the field theoretical model is effective to understand the fundamental properties.  
\\\hspace*{4.mm}
When there is no medium that is the zero density 
the interaction of pion with nucleon attracts our interest particularly for the treatment of the scattering.
To proceed the calculation we need to select the type of the interaction among the pseudoscalar and the pseudovector.   
Although the latter is chosen in the present study there is no reason to use it as far as the comparison of the numerical results with the experimental data.
\\\hspace*{4.mm}
One of the characteristics of the pseudovector coupling is the derivative interaction and the interacting part of the Hamiltonian contains the non-covariant term.
When the perturbative expansion is performed it is shown that the term cansels with the part of the delta function attached to the pion propagator 
and the problem is dissolved.
The difficulty of the system is the divergences appearing in the nucleon propagator since the procedure of the counter terms by the mass and the field
does not attain to the elimination of the divergences contrary to the pseudoscalar coupling interaction.
\\\hspace*{4.mm}
In order to obtain the nucleon propagator in the pseudovector coupling interaction the non-perturbative treatment is applicable 
to the Green's functions which consist of the field operators under the equation of motion.
The identity has not been derived between the vertex function and the propagator analogous to the Ward-Takahashi (W-T) identity of the quantum electrodynamics.
Instead of the identity the pion-nucleon-nucleon ($\pi$-$N$-$N$) vertex is found to be accompanied by the non-perturbative term.
\\\hspace*{4.mm}
The extended vertex is useful to construct the self-energy of nucleon. 
The divergence of the higher-order is cancelled fractionally and the result is convergent 
by dropping the zeroth and the first order terms of the perturbative expansion.
The lowest-order approximation is applied to the calculation of the magnetic moment of nucleon. 
It makes us possible to explain the experimental value 
supposing that the strength of the coupling constant of the pseudovector interaction decreases effectively. 
\\\hspace*{4.mm}
Recently the effect of the non-perturbative term on the process related to pion is examined 
in which the form factor of the $\gamma$-$\pi$-$\pi$ vertex is constructed by the nucleon loop diagram.
It has been found that the anomalous term of the electromagnetic interaction is indispensable to obtain the result compared with the experimental data.
Furthermore the self-energy is possibly changed from the value for the magnetic moment to the one appropriate to the loop diagram.  
Actually the idea is supported by the calculation of the pion-nucleon elastic scattering so as to give the phase-shift parameters.
The correction to the axial-vector form factor is investigated in the present study.
\\\hspace*{4.mm}
\section*{\normalsize{2 \quad The calculation of the axial-vector form factor}}
\hspace{4.mm}
The structure dependent part of the process $\pi^{+} \rightarrow \gamma + e^{+} + \nu_e$ is expressed by two form factors $F_V$ and $F_A$ \cite{Bryman}.
The former is associated with the vector part of the weak current and it has the property of the conservation of the electromagnetic current
ascribed to the structure of the loop integral.
For the latter part the electromagnetic current does not conserved singly in the emission of photon with the momentum $q$ as shown below. 
\\\hspace*{4.mm}
The loop of the structure dependent part $\Pi^A_{\mu\rho}$ is given as follows
\begin{eqnarray}
\Pi^A_{\mu\rho} \equiv {\rm Tr}\,[\,G(k) \gamma_5 \gamma_\rho G(p^\prime+k) \Gamma_\mu G(p+k) \gamma_5 \,]
\end{eqnarray}
\begin{eqnarray}
q^\mu \, \Pi^A_{\mu\rho} = -{\rm Tr}\,[\, \gamma_\rho G(p+k) G(-k) \,] + {\rm Tr}\,[\, \gamma_\rho G(p^\prime+k) G(-k) \,] \neq 0
\end{eqnarray}
in terms of the nucleon propagator $G$ where the momenta in the arguments are that of the pion $p$, the sum of the positron and the neutrino $p^\prime\,(=p-q)$
and the variable of the integral $k$.   
The W-T identity for the $\gamma$-$N$-$N$ vertex is used to obtain Eq. (2). 
The relation is exact and independent of the perturbative expansion.
\\\hspace*{4.mm}
It is possible that the photon is emitted from the incident pion. 
The relation of the current in the process is given by
\begin{eqnarray}
q^\mu \, \Pi_{\mu\rho} = -{\rm Tr}\,[\, \gamma_\rho G(p^\prime+k) G(-k) \,]
\end{eqnarray}
using the identity for the $\gamma$-$\pi$-$\pi$ vertex $\Gamma(p_f,p_i)$ along with the pion propagator $\it\Delta(p)$ 
\begin{eqnarray}
(p_f-p_i) \cdot \Gamma(p_f,p_i) = {\it\Delta(p_f)}^{-1} - {\it\Delta(p_i)}^{-1}
\end{eqnarray}
applied to the external line of pion between the initial ($p_i$) and the final ($p_f$) momenta.
Eq. (4) is analogous to the W-T identity for the $\gamma$-$N$-$N$ vertex and derived similarly by the procedure of the equation of motion.
\\\hspace*{4.mm}
In addition to them the emission of photon takes place at the external line of the positron in the final state.
Attaching the $\gamma$-positron-positron vertex $\Gamma_e^\mu$ and using the relation between the vertex and the positron propagator $G_e$  
proper to the on-shell positron 
\begin{eqnarray}
q \cdot \Gamma_e(p_e,p_e+q) G_e(p_e +q) \rightarrow 1
\end{eqnarray}
gives rise to the relation
\begin{eqnarray}
q^\mu \, \Pi^e_{\mu\rho} = {\rm Tr}\,[\, \gamma_\rho G(p+k) G(-k) \,]
\end{eqnarray}
These three relations are added and the current conservation 
\begin{eqnarray}
q^\mu \, (\Pi^A_{\mu\rho}+\Pi_{\mu\rho}+\Pi^e_{\mu\rho}) = 0
\end{eqnarray}
is verified.
It is the consequence of the gauge invariance in the quantum electrodynamics.
\\\hspace*{4.mm}
Among three parts the structure dependent part $\Pi^A_{\mu\rho}$ is related to the axial-vector form factor $F_A$ and meaningful to study the expression in detail.
While the current conservation is not preserved ($q^\mu \, \Pi^A_{\mu\rho} \neq 0$) 
the vertex $\Gamma_\mu$ still contains the part $\sim \sigma_{\mu\nu} q^\nu$ which definitely entails the conservation.
Then the anomalous interaction is essential to construct $F_A$ different from the case of $F_V$ 
where the point interaction ($\sim \gamma_\mu$) already has the property of the conservation.
The vertex of the electromagnetic interaction is approximated to $\Gamma_\mu \approx - \kappa i \sigma_{\mu\nu}q^\nu /2 M$ 
to proceed the perturbative calculation below.
The contribution of the point interaction is examined in the next section to make sure that it contains the part of the current conservation.
\\\hspace*{4.mm}
The conserved part of $\Pi^A_{\mu\rho}$ generally consists of a linear combination of two terms $\sim g_{\mu\rho} \, p  \cdot  q - p_\mu q_\rho$ 
and $\sim g_{\mu\rho} \, q^2 - q_\mu q_\rho$.
The form factor $F_A$ is related to the loop integral $\hat{T}_{\mu\rho}(p^\prime,p)$ as 
\begin{eqnarray}
- \sqrt{2} m \, G \, \hat{T}_{\mu\rho}(p^\prime,p) = F_A \, ( p  \cdot  q \, g_{\mu\rho} - p_\mu q_\rho) 
\end{eqnarray}
\begin{eqnarray}
\hat{T}_{\mu\rho}(p^\prime,p) \equiv \int \frac{d^4 k}{i (2 \pi)^4} \frac{T_{\mu\rho}(k,p^\prime,p)}{(k^2-M^2)((p^\prime+k)^2-M^2)((p+k)^2-M^2)}
\end{eqnarray}
\begin{eqnarray}
T_{\mu\rho}(k,p^\prime,p)\equiv {\rm Tr}[(\gamma\!\cdot\! k+M) \gamma_5 \gamma_\rho (\gamma\!\cdot\! (p^\prime+k)+M) \Gamma_\mu (\gamma\!\cdot\! (p+k)+M) \gamma_5 ]
\end{eqnarray}
in which $G \equiv 2 M f / m$ is the strength of the $\pi$-$N$-$N$ vertex and $f$, $m$ and $M$ are the pseudovector coupling constant,
the pion and the nucleon masses respectively.
\\\hspace*{4.mm}
The trace in Eq. (10) is calculated by using the procedure of the gamma matrix as follows
\begin{eqnarray}
T_{\mu\rho}(k,p^\prime,p) = 2 M^{-1} \kappa q^\nu [\, g_{\mu\rho}( M^2 p_\nu +(k^2-M^2)(p^\prime+k)_\nu 
+ \, p\cdot k \, p^\prime_\nu -p\cdot p^\prime k_\nu\nonumber
\end{eqnarray}
\begin{eqnarray}
 +(p^\prime+k)\cdot k \, p_\nu ) 
+(p^\prime +k)_\mu (k_\nu p_\rho - p_\nu k_\rho) +p_\mu k_\nu (p^\prime+k)_\rho -(\mu\leftrightarrow\nu) \,]
\end{eqnarray}
with the coefficient $\kappa=\kappa_p$ in the case of the $\gamma$-$p$-$p$ vertex. 
$T_{\mu\rho}(k,p^\prime,p)$ contains the terms giving infinity and in addition to the relation of the integral $\hat{k}_{\mu\nu}$ \cite{Kinpara}
the following type is needed
\begin{eqnarray}
&&\int \frac{d^4 k}{i (2 \pi)^4} \frac{k_\mu}{((p^\prime+k)^2-M^2)((p+k)^2-M^2)}
= -\frac{1}{2 (4 \pi)^2} P_\mu \,{\rm log} \frac{{\it\Lambda}^2}{M^2} \nonumber\\\nonumber\\
&&\, + \frac{1}{(4 \pi)^2} \int_0^1 d z (p^\prime_\mu z +p_\mu (1-z)) \,{\rm log} [\, 1-\frac{q^2}{M^2}z (1-z) \,] 
\end{eqnarray}
where $P \equiv p+p^\prime$.
The procedure of the regularization is used in which the cut-off parameter $\it\Lambda$ introduced is similar to our previous study \cite{Kinpara}.
\\\hspace*{4.mm}
The $k$-integral in Eq. (9) is given by
\begin{eqnarray}
\hat{T}_{\mu\rho}(p^\prime,p) = \frac{\kappa}{(4 \pi)^2 M}[\,(g_{\mu\rho}p\cdot q -p_\mu q_\rho) (-\frac{3}{2}+{\rm log} \frac{{\it\Lambda}^2}{M^2})\nonumber\\
-(g_{\mu\rho}q^2 -q_\mu q_\rho) \, {\rm log} \frac{{\it\Lambda}^2}{M^2}\,]
\end{eqnarray}
or using the relation $q = p-p^\prime$
\begin{eqnarray}
\hat{T}_{\mu\rho}(p^\prime,p) = \frac{\kappa}{(4 \pi)^2 M}[\,-\frac{3}{2} (g_{\mu\rho}p\cdot q -p_\mu q_\rho) 
+(g_{\mu\rho}p^\prime\cdot q -p^\prime_\mu q_\rho) \,{\rm log} \frac{{\it\Lambda}^2}{M^2}\,]
\end{eqnarray}
leaving terms which consist of the product of two momenta such as $\sim p q$ and $\sim q q$.
It is correct to drop the term $\sim g_{\mu\rho}q^2 -q_\mu q_\rho$ in Eq. (13) when the specific value $\it\Lambda \sim M$ is given to construct $F_A$.
The value of $\it\Lambda$ has already been used in the case of $F_V$ without the explanation for choosing it \cite{Kinpara}.
\\\hspace*{4.mm}
From Eqs. (8) and (13) the axial-vector form factor $F_A$ is obtained as
\begin{eqnarray}
F_A = \frac{3 \sqrt{2} f}{(4 \pi)^2} \,(\kappa_p-\kappa_n)(1-\frac{2}{3} \, {\log}\frac{{\it\Lambda}^2}{M^2}) 
\end{eqnarray}
where $\kappa = \kappa_p$ is used.
The process of the $\gamma$-$n$-$n$ vertex is the reversed direction of the loop for the $\gamma$-$p$-$p$ vertex and these are converted 
by the replacement of the momenta $p \rightarrow -p$ and $p^\prime \rightarrow -p^\prime$.
The minus sign in front of $\kappa_n$ is attributed to the trace in which the part of the axial-vector current contains $\gamma_5$.
Consequently $F_A$ is not symmetric under the transformation of $\kappa_p \leftrightarrow \kappa_n$.
\\\hspace*{4.mm}
\section*{\normalsize{3 \quad The non-perturbative term for the axial-vector form factor }}
\hspace{4.mm}
While the strength of the anomalous interaction is reduced by the cancellation ($\sim \kappa_p+\kappa_n$) in $F_V$ 
the minus sign due to the $\gamma_5$ matrix makes the value of $F_A$ increase ($\sim \kappa_p-\kappa_n$).
Then it is interesting to investigate the effect of the non-perturbative term. 
The procedure of the calculation for the loop diagram is explained by giving a few of the equations prior to examining the effect of the correction.
\\\hspace*{4.mm}
According to the non-perturbative relation the part of the $\pi$-$N$-$N$ vertex $\Gamma_0(p+k,k)$ defined by the perturbative expansion 
is certainly followed by the non-perturbative term
\begin{eqnarray}
\Gamma(p+k,k) = \Gamma_0(p+k,k) + G(p+k)^{-1} \gamma_5 + \gamma_5 G(k)^{-1} 
\end{eqnarray}
in terms of the nucleon propagator $G(p) = (\gamma\cdot p -M -\Sigma(p))^{-1}$ without referring to the isospin for the sake of simplicity.
The perturbative higher-order corrections are neglected ($\Gamma_0(p+k,k) \approx \gamma_5 \gamma\cdot p$) 
to replace the vertex $\Gamma(p+k,k)$ with the approximate form 
\begin{eqnarray}
\Gamma(p+k,k) \approx -2 M \gamma_5 - \Sigma(p+k) \gamma_5 -\gamma_5 \Sigma(k)
\end{eqnarray}
The first term is equivalent to the vertex of the pseudoscalar coupling interaction and it is the basic element to construct the loop diagram.  
The correction of the second and the third terms to the pseudoscalar model is our interest in the present study. 
\\\hspace*{4.mm}
Concerning the respective momenta of the nucleon propagators in the non-perturbative term 
they are assumed to have roughly the values satisfying the on-shell condition.
Then the self-energy is represented by a parameter $c \,(\equiv c_1^{(0)} = c_2^{(0)})$ which characterizes the process \cite{Kinpara}.
The value is determined by the method of the matrix inversion so as to reproduce the phase-shift in the pion-nucleon elastic scattering.
Consequently the $\pi$-$N$-$N$ vertex with the incident pion momentum $p$ is changed as
\begin{eqnarray}
\gamma_5 \rightarrow (1+c) \gamma_5 + \frac{c}{2 M} \gamma_5 \gamma\cdot p
\end{eqnarray} 
It is applied to the calculation of the trace in Eq. (10).
The procedure of the non-perturbative term to correct the loop integral has been performed also in the calculation of the vector form factor $F_V$ \cite{Kinpara}.
\\\hspace*{4.mm}
Taking account of the vertex in Eq. (18) the numerator $T_{\mu\rho}(k,p^\prime,p)$ is modified as 
\begin{eqnarray}
T_{\mu\rho}(k,p^\prime,p) \rightarrow (1+c) \, T_{\mu\rho}(k,p^\prime,p) + \frac{c}{2} \, T^{\,\prime}_{\mu\rho}(k,p^\prime,p)
\end{eqnarray}
\begin{eqnarray}
T^{\,\prime}_{\mu\rho}(k,p^\prime,p) = 2 M^{-1} \kappa q^\nu [\, -g_{\mu\rho} (M^2+3k^2) p_\nu  \nonumber\\\nonumber\\
+4 \,(g_{\mu\rho} p_\sigma k_\nu k^\sigma +k_\mu k_\rho p_\nu) -(\mu\leftrightarrow\nu) \,]
\end{eqnarray}
where the terms of the zeroth and the first order in $k$ giving only minor effects are neglected.
Using the above relation with $c \neq 0$ it yields  
\begin{eqnarray}
F_A \rightarrow (1+c) \, F_A - \frac{2 \sqrt{2} f}{(4 \pi)^2} (\kappa_p-\kappa_n) (1-{\rm log}\frac{{\it\Lambda}^2}{M^2}) \, c 
\end{eqnarray}
where the $F_A$ is given in Eq. (15).
\\\hspace*{4.mm}
In our previous study two parameters $\it\Lambda$ and $c$ are adjusted to reproduce the experimental data of $F_V$ and the radius of pion $r_\pi$ simultaneously. 
The way to determine these parameters is not taken here 
since the assumed form of the structure dependent term constrains to use $\it\Lambda \sim M$ in Eqs. (13) and (14).
Meanwhile the parameter $c$ is obtained by the lowest-order approximation of the expansion for the self-energy 
and which is suitable for the calculation of the anomalous magnetic moment.
But the application of the value to the form factors of pion has revealed that the numerical results are much larger than the respective experimental values.
\\\hspace*{4.mm}
As mentioned above the calculation of the pion-nucleon elastic scattering is employed to reduce the $\kappa_p$ by using the smaller value $c=-0.39$.
It improves the numerical value of $F_A$ as well as the case of $F_V$. 
Another coefficient $\kappa_n$ is independent of the self-energy term in the W-T identity and remains intact.
Using the value of $c$ for the relation in Eq. (21) with $\it\Lambda = M$ and $f = 1$ the calculation results in $F_A = 0.0309$.
It is much larger than the value of the experiment $F_A = 0.0116\pm0.0016$ \cite{PDG} 
and our interest is to proceed the calculation of the terms generated by the point interaction.
\\\hspace*{4.mm}
While the gauge invariance gives the conservation of the current as a whole the structure dependent part does not satisfy the condition independently.
Then we need to pick out the part which has the assumed form and examine the effect to $F_A$.
Using the loop $\Pi^A_{\mu\rho}$ in Eq. (1) the $k$-integral is 
\begin{eqnarray}
q^\mu \int \frac{d^4 k}{i(2 \pi)^4} \, \Pi^A_{\mu\rho} = - 4 M q_\rho \, J({p^\prime}^2) \, + 4 M p_\rho (J({p^\prime}^2)-J({p}^2))
\end{eqnarray}
\begin{eqnarray}
J(x) = \frac{1}{(4 \pi)^2} {\rm log} \frac{{\it\Lambda}^2}{M^2} + \frac{x}{6 (4 \pi)^2 M^2} +O(x^2)
\end{eqnarray}
where $J(x)$ is expanded in the series of $x$ up to the $\sim x$ order.
\\\hspace*{4.mm}
Removing $q^\mu$ from Eq. (22) the following relation
\begin{eqnarray}
\hat{\Pi}^A_{\mu\rho} \equiv \int \frac{d^4 k}{i(2 \pi)^4} \, \Pi^A_{\mu\rho} 
= \frac{2}{3 (4 \pi)^2 M} (\, - q^2 g_{\mu\rho} - P_\mu p_\rho + 2 p_\mu q_\rho \,)\nonumber\\ \nonumber\\
 -4 M J(p^2) g_{\mu\rho}\, +{\it\Gamma}_{\mu\rho}  
\end{eqnarray}
\begin{eqnarray}
{\it\Gamma}_{\mu\rho} = -\frac{2}{3 (4 \pi)^2 M}( p \cdot q \, g_{\mu\rho} - p_\mu q_\rho )
\end{eqnarray}
is obtained. 
In order to determine the form of ${\it\Gamma}_{\mu\rho}$ obeying the relation $q^\mu {\it\Gamma}_{\mu\rho} = 0$ 
the loop diagram has been calculated using the point interaction $\Gamma_\mu = \gamma_\mu$ in Eq. (1).
When the condition $\it\Lambda = M$ is applied the constant term ($\sim J(p^2) g_{\mu\rho}$) does not vanish completely.
To eliminate the term the value of $\it\Lambda$ is changed from $M$ to $\it\Lambda = M( {\rm 1} - m^{\rm 2} /{\rm 12} M^{\rm 2} )$ 
due to the condition $J(p^2) = 0$.
The shift of $\it\Lambda$ does not make remarkable effects on the numerical values of the form factors 
and we need to include the terms of the order $\sim m^2/M^2$ which are not considered in the present study.
\\\hspace*{4.mm}
Following the extension of the vertex in Eq. (18) the non-perturbative term is taken into account and $\hat{\Pi}^A_{\mu\rho}$ in Eq. (24) is changed to
\begin{eqnarray}
\hat{\Pi}^A_{\mu\rho} \rightarrow (1+c) \, \hat{\Pi}^A_{\mu\rho} + \frac{c}{2} \, \Pi^\prime_{\mu\rho}
\end{eqnarray}
\begin{eqnarray}
\Pi^\prime_{\mu\rho} = \frac{4}{3 (4 \pi)^2 M} \,[\,( 1 - \, {\rm log}\frac{{\it\Lambda}^2}{M^2} )\, P_\mu p_\rho 
-(\frac{1}{2} + \, {\rm log}\frac{{\it\Lambda}^2}{M^2}) p_\mu q_\rho         \nonumber\\
+p_\mu p_\rho \, {\rm log}\frac{{\it\Lambda}^2}{M^2} + ( \frac{1}{4} + {\rm log}\frac{{\it\Lambda}^2}{M^2} ) g_{\mu\rho} p^2 \,]
+{\it\Gamma}^\prime_{\mu\rho}
\end{eqnarray}
\begin{eqnarray}
{\it\Gamma}^\prime_{\mu\rho} = \frac{10}{3(4 \pi)^2 M} (1 - \frac{4}{5} \, {\rm log}\frac{{\it\Lambda}^2}{M^2} ) ( p \cdot q \, g_{\mu\rho} - p_\mu q_\rho )
\end{eqnarray}
The ${\it\Gamma}^\prime_{\mu\rho}$ is the counterpart of ${\it\Gamma}_{\mu\rho}$ in Eq. (25)
and also determined by the calculation of the loop diagram.
Remaining only these terms the correction to the structure dependent part is
\begin{eqnarray}
F_A = \frac{4 \sqrt{2} f}{3 (4 \pi)^2} [\, 1+c-\frac{5}{2} \, (1 - \frac{4}{5} \, {\rm log}\frac{{\it\Lambda}^2}{M^2} ) \, c \,]
\end{eqnarray}
Adding Eqs. (21) and (29) it yields the numerical value $F_A \,$=0.0498.
\\\hspace*{4.mm}
It is surprising that the non-conserving part of $\Pi^\prime_{\mu\rho}$ is different between the direct calculation of the loop diagram with the point interaction
and the calculation of the quantity for $q^\mu \Pi^\prime_{\mu\rho}$ using the identity in the free propagator of nucleon.
The former way is employed to proceed the investigation for the moment.
The property is not seen in the calculation of $\hat{\Pi}^A_{\mu\rho}$.
\\\hspace*{4.mm}
When the photon emitted is not real ($q^2$$\neq$ 0) the right side of Eq. (8) is extended as  
\begin{eqnarray}
F_A \, (( p  \cdot  q - q^2 ) g_{\mu\rho}  - p_\mu q_\rho) + R \, q^2 g_{\mu\rho}
\end{eqnarray}
in terms of the second axial-vector form factor $R$ \cite{PDG}.
The calculation yields the numerical value $R$=0.0570 in agreement with the experimental value $R$=0.059$\pm{}_{0.008}^{0.009}$ \cite{PDG}.
Then the present approach of the point interaction is suitable for the description of the breaking term of the current conservation $R$ 
although the inclusion of the conserved part prevents the result of $F_A$ from approaching the experimental data. 
\\\hspace*{4.mm}
It is required to connect $R$ with the radius of pion $r_\pi$ 
by virtue of the approximate relation $R = m f_\pi r_\pi^2 /3$ along with the decay constant $f_\pi$.
In fact it has been verified that the calculation of the loop diagram gives the value of $r_\pi$ well.
Nevertheless, the result of the value $F_A$ is much larger than that of the experiment and would probably need the higher-order corrections
or the extension of the present model of the self-energy to account for the decay process quantitatively.
\\
\section*{\normalsize{4 \quad Summary and remarks }}
\hspace*{4.mm}
While the calculation of the vector form factor does not possess the means to determine the value of the cut-off 
the assumed form of the axial-vector form factor enables to find it in the vicinity of the nucleon mass.
Therefore it is possible to obtain the convergent result for each process which consists of the nucleon loop. 
The numerical values of the form factors of pion except for the axial-vector form factor are consistent with the experimental data. 
It indicates that the parameter for the self-energy in the non-perturbative term is reduced from the value suitable for the description of the magnetic moment.
At this time there is no explanation for the reason why two different values are coexistent and the relation between them.
\hspace{4.mm}

\end{document}